
\magnification = 1200
\baselineskip=16pt
\null\vskip-30pt
\rightline{SWAT/95/82}
\rightline{hep-lat/9509072}
\rightline{September 1995}
\vskip0.5 truecm
\centerline{\bf ELECTROMAGNETIC SELF-DUALITY IN A LATTICE MODEL}
\vskip 0.5 truecm
\centerline{{\bf Simon Hands}}
\centerline{\it Department of Physics,}
\centerline{\it University of Wales, Swansea,}
\centerline{\it Singleton Park,}
\centerline{\it Swansea SA2 8PP, U.K.}
\vskip 0.5 truecm
\centerline{and}
\vskip 0.5 truecm
\centerline{\bf John B. Kogut}
\centerline{\it Department of Physics,}
\centerline{\it University of Illinois at Urbana-Champaign,}
\centerline{\it 1110 West Green Street,}
\centerline{\it Urbana, IL 61801-3080, U.S.A.}
\vskip 1.5 truecm
\centerline{\bf Abstract}

{\narrower
\noindent
We formulate a Euclidean lattice theory of interacting elementary
spin-half electric
and magnetic charges, which we refer to as electrons and magnetic monopoles
respectively. The model uses the polymer representation of the fermion
determinant, and exhibits a self-dual symmetry provided electric charge $e$
and magnetic charge $g$ obey the minimal Dirac quantisation condition
$eg=2\pi$. In a hopping parameter expansion at lowest order,
we show that virtual
electron and monopole loops contribute radiative corrections of opposite
sign to the photon propagator. We argue that in the limit $e\to0$, fermion mass
$\mu\to0$,
the model describes QED together
with strongly interacting monopoles whose chiral symmetry is spontaneously
broken. Prospects for the existence of an interacting continuum limit at
the self-dual point $e=g$ are discussed.
\smallskip}


\vfill\eject
\noindent{\bf1. Introduction}

Ever since Maxwell's equations were written down, it has been natural to
speculate why there exist fundamental electric charges but apparently no
magnetically-charged counterparts. At the classical level, it is
straightforward
to modify Maxwell's equations for electric and magnetic field strengths (in
appropriate units) by the
introduction of magnetic charge and current densities
$(\rho_m,{\bf j}_m)$ analogous to the electric quantities $(\rho_e,{\bf j}_e)$:
$$\eqalign{
\partial_0{\bf E}&=\nabla\wedge{\bf B}-{\bf j}_e,\cr
\partial_0{\bf B}&=-\nabla\wedge{\bf E}-{\bf j}_m,\cr
\nabla.{\bf E}&=\rho_e,\cr
\nabla.{\bf B}&=\rho_m.\cr}\eqno(1.1)$$
The system of equations (1.1) is invariant under a duality transformation:
$$\eqalign{
{\bf E}\mapsto{\bf B}\;\;\;&;\;\;\;{\bf B}\mapsto-{\bf E}\cr
(\rho_e,{\bf j}_e)\mapsto(\rho_m,{\bf j}_m)\;\;\;&;\;\;\;
(\rho_m,{\bf j}_m)\mapsto-(\rho_e,{\bf j}_e).\cr}\eqno(1.2)$$
The cost of this modification is that it is no longer possible to describe
the electromagnetic field in terms of an abelian vector potential $A_\mu$ which
is both
globally defined and non-singular. This raises difficulties, because the
standard
quantum-mechanical description of electromagnetism is formulated in terms
of the potential rather than the field strengths, and indeed the success
of the prototype
quantum field theory, QED, depends critically on the local gauge symmetry
transformations which relate physically equivalent configurations of the
$A_\mu$ field.
Dirac [1] showed how to introduce
magnetic monopoles as pointlike sources of magnetic flux by connecting them
to line singularities in $A_\mu$; so-called Dirac strings. Despite the
singularity the string
has no observable effect on a particle carrying electric charge $e$,
in either classical or quantum mechanics,
if the strength of the magnetic charge $g$ is quantized according
to the famous condition
$$eg=2n\pi\hbar,\eqno(1.3)$$
where $n$ is an integer. Planck's constant will henceforth be set to one.
The spatial position of the string can be
changed under gauge transformations, but the magnetic flux emanating from the
monopole is uneffected. The Dirac quantization condition (1.3), or
variations of it, has been central to all subsequent attempts to create
a quantum theory which includes monopoles (for a comprehensive review
see [2]).

 Attempts to construct a quantum
field theory containing both particles with electric charge and
particles with magnetic charge -- so-called Quantum
ElectroMagnetoDynamics (QEMD) -- began with Cabibbo and Ferrari [3] and
Schwinger [4]. The latter formulation
is in the Hamiltonian framework, and contains fermionic electron and
monopole fields each interacting with its own vector potential field.
The two potentials are not independent. The model also has
an explicit dependence on
a vector {\bf s} defining the direction of a Dirac-type singularity,
needed to relate the potentials to the current distributions defined by the
matter fields. Later an equivalent Lagrangian formulation was given by
Zwanziger [5]. In both cases the quantization condition was modified to
be
$$eg=4n\pi.\eqno(1.4)$$
The string-dependence obscures the Lorentz invariance of the
results. In addition, since the string's position is gauge-dependent,
insistence on a fixed string vector $s_\mu$ leaves no possibility of
a gauge invariant formalism. An alternative approach, beginning with
[3], allows the string, rather than being fixed and infinite, instead to
run between monopole - antimonopole pairs, resulting in a description
in terms of closed particle world-lines spanned by surfaces known as
Dirac sheets. This formulation is gauge invariant and stresses the
path-dependence of the particle dynamics -- and hence is inherently
non-local. The requirement of surface-independence once again results in
the Dirac quantization condition. The world-line approach was used by Brandt,
Neri and Zwanziger [6] to demonstrate that gauge-invariant Green functions
in the Lagrangian formulation are independent of the vector $s_\mu$ defining
the string.

As we have sketched, a common feature of the various formulations, some of
which
have been shown to be equivalent [2], has been the impossibility of
constructing a manifestly local, covariant and gauge invariant formulation.
Another problematic issue has been how to proceed once the model is
formulated -- various authors have succeeded in finding ``Feynman rules'',
but these are of limited use due to the condition (1.3): QEMD is strongly
interacting. Indeed, the origin of the quantization condition,
the requirement that $\exp(ieg)=1$, is inherently non-perturbative.

Calucci and Jengo [7], realising that the strongly-interacting nature of
the problem necessitates new techniques, regularised a version of the
worldline formalism [8] on a Euclidean spacetime lattice. A
non-perturbative regularisation must be used if the ultraviolet behaviour
of the model is to be discussed adequately. Their model has two field strength
tensors, defined on the direct and dual lattices respectively,
and requires the more restrictive quantisation condition (1.4). The
lattice is used as little more than a formal framework, but enables the use
of a proper time technique to calculate the effects of virtual particle loops
on the interaction among electric and magnetic charge strengths. Their claim
is that if $e_0$ and $g_0$ are ``bare'' charge strengths satisfying (1.4),
then the ``renormalised'' charges $e_R$, $g_R$ satisfy
$$e_Rg_R=e_0g_0,\eqno(1.5)$$
ie, $eg$ is a renormalisation group invariant. The result follows essentially,
as we shall see, because of an extra factor of $i$ in the electron -- monopole
interaction which reverses the sign of the vacuum polarisation diagram:
monopoles ``anti-screen'' electric charge, and vice-versa. Although
persuasive, this result is in contradiction to that obtained by other
authors [9]. Moreover, the Feynman rules developed by Calucci and Jengo
contain an ordering ambiguity which requires an artificial distinction to
be drawn between valence and virtual current loops, once again reflecting
the non-perturbative nature of the quantisation condition and suggesting that
any approach based on series expansions in $e$ and $g$ will not be fully
satisfactory.

As pointed out in [7], the self-dual point $e=g$ appears to define a fixed
point of the renormalisation group, since if the duality symmetry is respected
by the regularisation then $e_R=g_R$ and $e_Rg_R={\rm constant}$ together
imply that neither $e$ nor $g$ is renormalised. The existence of an
ultraviolet fixed point, or continuum limit, for abelian theories such as
QEMD or indeed QED has been a matter of speculation for many years [10]. In
perturbative QED we know the relation between bare and renormalised electric
charge:
$$e^2_R(\mu)={{e^2_0(\Lambda)}\over{1+{\displaystyle{e_0^2\over{6\pi^2}}}
\ln{\displaystyle \left({\Lambda
\over\mu}\right)}}},\eqno(1.6)$$
where $\Lambda$ is the cutoff and $\mu$ a physical scale, implying that
$e_R\to0$ as $\Lambda/\mu\to\infty$. This phenomenon is known as triviality
-- QED can only exist as an interacting theory if the cutoff is finite. An
alternative scenario requires the existence of a zero in the $\beta$-function
describing renormalisation group flow of the interaction strength beyond
the perturbative regime. One possibility was advanced by Miranskii and
co-workers, who investigated the self-energy of the electron using a truncated
system of Schwinger-Dyson equations [11]. For a critical value
$e_c=2\pi/\surd3$ the chiral symmetry protecting the electron from acquiring
mass is spontaneously broken. A physically meaningful ground state can only
be found if the bare charge $e_0$ obeys
$${e_0^2(\Lambda)\over e_c^2}=1+\left({\pi\over\ln(\Lambda/\mu)}\right)^2,
\eqno(1.7)$$
where $\mu$ is now related to the electron mass in the broken phase.
Hence the continuum limit is taken as $e_0\to e_c$.
Note that $e_c$ is greater than either of the ``minimal'' self-dual values
in QEMD $e=\sqrt{2\pi},\sqrt{4\pi}$.

The Miranskii fixed-point hypothesis
received support from numerical simulations of non-compact lattice QED
(NCQED) [12],
where it was found that for sufficiently strong electric charge the model
exhibits a continuous phase transition from the perturbative phase to one
in which chiral symmetry is broken, signalled by the non-vanishing of the
condensate $\langle\bar\psi\psi\rangle$. However, intensive computer simulation
has not found the scaling behaviour consistent with Miranskii's prediction
of an essential singularity as $e\to e_{c+}$: although there has been some
dispute in the literature, currently the existing numerical data for
the equation of state relating $\langle\bar\psi\psi\rangle$, $e$
and the bare electron mass $\mu$ are best
fitted by a critical coupling $e_c$ in the region 2.2-2.3 (for the
case of four dynamical fermion species), that is, less than the value
predicted by the Scwinger-Dyson approach, with a
power-law singularity with either non-mean field exponents [13], or mean
field exponents with logarithmic corrections [14]. In either case the
the continuum limit is thought to consist of both electrons and
bosonic $e^+e^-$ bound states. The mean field fit is motivated
by a picture in which the bound state
charge radii and physical interaction strengths vanish so
that the model is trivial.

To a large extent, the conclusions drawn from
numerical simulations are subject to theoretical prejudice; the true fixed
point theory, if it exists, may contain many other interactions apart from
the electron-photon interaction of the minimal gauge coupling (alternatives
for the lattice electron-photon interaction have been explored in [15,16]).
Constraints on interaction Lagrangians
due to the requirement of perturbative renormalisability may not
apply at a strongly-coupled fixed point. Operators such as the
chirally-symmetric $(\bar\psi\psi)^2-(\bar\psi\gamma_5\psi)^2$ may acquire
large anomalous scaling dimensions through non-perturbative effects and
hence become relevant at the fixed point [17]. It is difficult to
know {\it a priori\/} how to restrict the space of bare theories which need
to be explored before an interacting continuum limit can be excluded. A
defensible viewpoint is that the mere existence of an observed regime where
perturbative QED applies (as an effective theory) implies the existence
of an underlying theory containing massive degrees of freedom and/or
short-ranged interactions which are so far imperfectly determined.
Of course, it is
more usual to consider electromagnetism as part of a unified description
of fundamental interactions, though it is worth recalling that the Standard
Model gauge group contains U(1) as a subgroup,
so the issue of a continuum limit
for abelian gauge theories cannot as yet be ignored.

In this paper we shall explore the notion that the bare theory, and
hence the fixed-point theory, are constrained to have the symmetry (1.2)
of electromagnetic duality, by modifying lattice QED so that it has both
electron and fermionic magnetic monopole degrees of freedom; in other words,
we construct a lattice QEMD. We have been motivated in part by the observation
that at the large values of $e$ in the vicinity of the observed chiral
transition, the lattice photon field may be interpreted as containing line
dislocations defined on the dual lattice, which are akin to magnetic
currents [18]. The monopole dynamics in NCQED are
unconventional, however, since the Dirac sheet in this formulation has
non-vanishing action density, which means that in effect the monopoles
probably exist in tightly bound dipole pairs with no associated long-ranged
electromagnetic fields.
Lattice monopoles bear a closer resemblance
to Dirac ones when they occur as excitations in compact or periodic abelian
gauge theory [19,20], in which the Dirac sheet costs no action and is
physically unobservable; in this case the monopoles act as sources for
long-ranged Coulomb fields. Banks {\it et al\/} [19] showed how the partition
function for the Villain
form of U(1) lattice gauge theory may be represented as a Coulomb gas of
monopole point charges in three Euclidean dimensions or
monopole world line loops
in four dimensions by using an exact lattice duality transformation. By
combining this formulation of lattice monopoles with the ideas
of Calucci and Jengo, we are able to construct a theory with electron and
monopole currents interacting via a Coulombic potential, with the charge
strengths given by the less restrictive quantisation condition (1.3)
with $n=1$. Field strengths and gauge potentials, with all their associated
complications [2], play a secondary role.
The cost is that, as in all previous attempts, we are unable to
give a local covariant action. Instead, there is a partition function written
in terms of ``polymer configurations'' on both direct and dual lattices,
based on the polymer representation of the lattice fermion determinant
introduced by Karowski {\it et al} [21]. Because the fundamental excitations
are fermions, an additional penalty is that the terms in the partition
function fluctuate in sign.
In the limit $e\to0$, $g\to\infty$, we argue that the electron
and magnetic sectors decouple, and that the model reduces to a
local action describing electrons interacting via photon exchange as in
QED. The monopoles' chiral
symmetry is spontaneously broken, so that the only light degrees of freedom
in the monopole sector are monopole -- antimonopole
``magnetopions'' corresponding
to the Goldstone mode associated with the broken symmetry. The only
conceivable point where both electrons and monopoles can exist as light
degrees of freedom is the self-dual point, which is therefore a candidate
continuum limit for the model.

The rest of the paper is organised as follows. In Section 2 we introduce
lattice QEMD in a heuristic fashion, by first discussing non-compact lattice
QED in the polymer representation, and then reviewing the transformation
from Villain lattice model to the monopole gas, before finally writing
the QEMD partition function. We also briefly review connections with other
approaches to QEMD.
In section 3 we discuss the possible phase
structure of the model, using exact results, analytic approximations and known
numerical results. A hopping parameter expansion to lowest order is used to
find that electron and monopole loops induce radiative corrections of
opposite sign in the photon propagator, in agreement with [7].
Unfortunately it seems that in the minimal model there is no value of the
couplings, including the self-dual point,
for which both electrons and monopoles are light degrees of freedom;
therefore probably
no interacting continuum limit exists. The reasons
are discussed in the final section, along with
possible modifications of the model which
may improve the prospects for a continuum limit,
and phenomenological consequences.
\vskip 1.0 truecm
\noindent{\bf2. A Lattice QEMD}

In this section we build the QEMD partition function in simple stages.
We begin with the partition function of non-compact lattice QED (NCQED):
$$Z_{NCQED}=\int{\cal D}\theta\det(D{\!\!\!\! /}\,[\theta]+\mu)
\exp\left(-{1\over4}\sum_{x\mu\nu}\Theta_{\mu\nu}^2(x)\right),\eqno(2.1)$$
where the determinant results from integrating over fermion fields of
bare mass $\mu$, leaving
the photon fields $\theta_\mu\in(-\infty,+\infty)$ as real dynamical variables
defined on the links of a hypercubic lattice. The field strength tensor
$\Theta_{\mu\nu}$ is then defined by
$$\Theta_{\mu\nu}(x)\equiv\Delta^+_\mu\theta_\nu(x)-\Delta^+_\nu
\theta_\mu(x),\eqno(2.2)$$
with $\Delta^+_\mu$ the forward lattice difference operator. We choose to
use the staggered fermion formulation (this is not crucial in what
follows), so the fermion
kinetic operator is given by
$$D{\!\!\!\! /}\,(x,y)={1\over2}\sum_\mu\eta_\mu(x)\left[
\delta_{y,x+\hat\mu}\exp(ie\theta_\mu(x))
-\delta_{y,x-\hat\mu}\exp(-ie\theta_\mu(x-\hat\mu))\right],\eqno(2.3)$$
where the Kawamoto-Smit phases $\eta_\mu(x)\equiv(-1)^{x_1+\cdots+x_{\mu-1}}$.
The electric charge $e$ enters only via the gauge covariant connection
in the derivative.
To avoid a proliferation of summation signs in the formul\ae\/ which follow
we adopt the convention that repeated spacetime indices and repeated
spatial arguments are summed over, even in squared quantities.
We shall also employ the subscript $\ell$ in connection with oriented
variables associated with a particular link.

The partition function in (2.1) is based on an action with a local
gauge invariance, of course.
For $Z$ to be strictly well-defined the gauge action must be gauge
fixed to avoid the partition function from diverging on integration
over a flat direction.
We now immediately introduce one of the major conceptual tools we will need
-- the polymer representation of the fermion
determinant introduced by Karowski {\it et al\/} [21]. We write
$$Z=\int{\cal D}\theta \mu^N\left\{\sum_{\{{\cal C}\}}\kappa^{N_\ell({\cal C})}
{\rm phase}({\cal C},\theta)\right\}
\exp\left(-{1\over4}\Theta_{\mu\nu}^2(x)\right),\eqno(2.4)$$
where $\kappa$ is the hopping parameter $1/2\mu$ and ${\cal C}$ is a
partition of the lattice in which every site is
either isolated (a monomer), joined to a nearest neighbour by a double ``bond''
(a dimer), or visited once and once only by a closed oriented self-avoiding
loop of single bonds (a polymer). The set of polymers in a particular
configuration ${\cal C}$ will be denoted $\{\Gamma\}$.
Each configuration ${\cal C}$ corresponds to
a unique term in the expansion of the determinant: the $N_m$ monomers
correspond to diagonal elements, and hence contribute a factor $\mu^{N_m}$, and
the $N_\ell$  bonds correspond to off-diagonal hopping terms. Clearly we
have
$$N_m+N_\ell=N\eqno(2.5)$$
where $N$ is the lattice volume. The phase factor has a number of
different components: a signed permutation factor from the expansion of
the determinant; a sign from the path over forward and backward links
and the Kawamoto-Smit phases (we shall refer to this as the ``signed KS
phase''); and finally the gauge connection $\prod_{\ell\in{\cal
C}}\exp(ie\theta_\ell)$. A moment's thought reveals that monomers and
dimers give trivial contributions to the phase -- only the polymers
$\{\Gamma\}$ contribute. The smallest polymer which gives a
negative sign is a non-planar loop of length six. We can rewrite
$Z$ as follows:
$$\eqalign{Z&\propto\int{\cal D}\theta\sum_{\{{\cal
C}\}}\left[\kappa^{-N_m}
(-1)^{N_\Gamma}\left(\prod_{\ell\in\{\Gamma\}}\eta_\ell\right)
\left(\prod_{\Gamma\in\{\Gamma\}}
U(\Gamma)\right)\exp\left(-{1\over4}\Theta_{\mu\nu}^2(x)\right)
\right]\cr
&\equiv\sum_{\{{\cal C}\}}\kappa^{-N_m}(-1)^{N_\Gamma}
\left(\prod_{\ell\in\{\Gamma\}}\eta_\ell\right)
Z_{photon}[\Gamma],\cr}\eqno(2.6)$$
where $N_\Gamma$ is the number of polymers, $\eta_\ell$ is the signed KS
factor associated with the link $\ell$, and
$U(\Gamma)=\prod_{\ell\in\Gamma}\exp(ie\theta_\ell)$ is the Wilson
loop associated with the polymer $\Gamma$ (note that since the polymers
are oriented, then both $U(\Gamma)$ and $U^*(\Gamma)$ appear as separate
terms in $Z$ with equal weight). $Z_{photon}[\Gamma]$ is simply the partition
function for free photons coupled to a Wilson loop distribution
$\{\Gamma\}$. Note the sign factor associated with the number of loops:
the relation with Fermi-Dirac statistics is clear. Note also that
despite the superficial similarity this is not the conventional
hopping parameter
expansion -- the sum over $\{{\cal C}\}$ is finite on a finite system,
and this expression for $Z$ is valid for all $\mu$.

Now let's focus on $Z_{photon}[\Gamma]$. This can be written
$$Z_{photon}[\Gamma]=\int_{-\infty}^{+\infty}\prod_\ell d\theta_\ell
\exp\left(-{1\over4}(\Delta_\mu^+\theta_\nu(x)-\Delta_\nu^+\theta_\mu(x))^2
+ie\sum_{\ell\in\{\Gamma\}}j_\ell\theta_\ell\right),\eqno(2.7)$$
where $j_\ell$ is the characteristic function of the polymer
configuration, ie
$$j_\ell=\cases{+1,&if $\ell$ a forward link $\in\{\Gamma\}$;\cr
               -1,&if $\ell$ a backward link $\in\{\Gamma\}$;\cr
                0,&otherwise.\cr}\eqno(2.8)$$
Of course, $j_\ell$ is nothing other than the electric current. By
performing a shift of the integration variables $\theta_\ell$, we can
rewrite
$$Z_{photon}[j]=Z_{photon}[0]\exp\left(-{e^2\over2}
j_\mu(x)v_{\mu\nu}(x-y)j_\nu(y)\right).\eqno(2.9)$$
Here $v_{\mu\nu}(x)$ is the lattice Coulomb propagator, whose existence
requires a gauge-fixing term in the photon action. In Feynman gauge it
satisfies
$$\Delta_\rho^+\Delta_\rho^-v_{\mu\nu}(x)=-\delta_{\mu\nu}\delta(x),
\eqno(2.10)$$
where $\Delta_\mu^-$ is the
backward difference operator. We have written the arguments $x$, $y$ to
stress that the interaction between the current loops is non-local in
this representation. The lattice Coulomb propagator is finite for
zero spatial separation, and for large $\vert x\vert$ behaves as
$\vert x\vert^{-2}$. Hence, in the polymer representation of the fermion
determinant, lattice QED resembles a Coulomb gas of unit charged
non-intersecting loops of variable sign:
$$Z_{NCQED}\propto Z_{photon}[0]\sum_{\{{\cal
C}\}}\kappa^{-N_m}(-1)^{N_\Gamma}\left(\prod_{\ell\in\{\Gamma\}}\eta_\ell\right)
\exp\left(\sum_{\ell,\ell^\prime\in\{\Gamma\}}-{e^2\over2}j_\ell
v_{\ell\ell^\prime}j_{\ell^\prime}\right).\eqno(2.11)$$
It is important to note, however, that the role of monomers and dimers
is also important, as we shall see.

Next, we review another well-known
model -- the Villain approximation to U(1) lattice gauge theory [19].
$$Z_{Villain}=\int_{-\pi}^\pi\prod_{\ell}{{d\theta_\ell}\over{2\pi}}
\sum_{\{n\}}\exp\left({{i\over2}n_{\mu\nu}(x)\Theta_{\mu\nu}(x)}-{e^2\over4}
n_{\mu\nu}^2(x)\right),\eqno(2.12)$$
where $n_{\mu\nu}(x)$ are integers defined on each plaquette of the
lattice. The Villain
action looks a little strange, but is simply the
convolution of the Gaussian non-compact gauge action $\Theta^2$ with an
infinite
array of delta functions having period $2\pi$. Around any one of these
locations the Villain action has approximately the Gaussian form of
NCQED, but by construction it is also periodic in $\Theta$, and in this
respect is similar to the Wilson form $(1-\cos\Theta)$. Its great
virtue is the existence of a sequence of exact transformations
which mean its phase structure may be understood in terms of monopole
excitations.
The integral
over $\{\theta\}$ can be performed immediately to leave a constrained
action of the $n_{\mu\nu}$:
$$Z_{Villain}=\sum_{\{n\}}\delta(\Delta_\nu^-n_{\nu\mu})
\exp\left(-{e^2\over4}n_{\mu\nu}^2(x)\right).\eqno(2.13)$$
The constraint can be solved by rewriting in terms of another integer
variable $l$:
$$n_{\mu\nu}(x)=\epsilon_{\mu\nu\lambda\kappa}\Delta_\lambda^+
l_\kappa(\tilde x).
\eqno(2.14)$$
Here, $l$ is defined on the links of the {\it dual\/} lattice, as
indicated by the tilde on $x$,
and $\epsilon$ is the totally antisymmetric tensor. As a
rule, any expression containing an odd number of $\epsilon$ symbols
relates variables living on direct and dual lattices. Note that (2.14)
does not uniquely specify $l$, or even require it to be integer-valued,
since the definition of $n$ remains invariant under
$$l_\mu(\tilde x)
\mapsto l_\mu(\tilde x)+\Delta^+_\mu\Lambda(\tilde x),\eqno(2.15)$$
where $\Lambda$ is
an arbitrary scalar function.
Now we can write
$$Z_{Villain}=\sum_{\{l\}}\exp\left(-{e^2\over4}
(\Delta_\mu^+l_\nu(x)-\Delta_\nu^+l_\mu(x))^2\right).\eqno(2.16)$$
This expression is very reminiscent of NCQED, except that the dynamical
variables are integers, not real numbers. The partition function may
now be reexpressed, first by
using the Poisson formula
$$\sum_{\{l\}}f(l)=\sum_{\{m\}}\int_{-\infty}^\infty d\phi f(\phi)
e^{2\pi im\phi},\eqno(2.17)$$
where the $m$ are integer,
and then by shifting the $\phi$ variable in the resulting Gaussian
integral as before. The result is
$$Z_{Villain}=Z_{photon}\sum_{\{m\}}\exp\left(\sum_{\ell\ell^\prime}
-{g^2\over2}m_{\tilde\ell} v_{\tilde\ell\tilde\ell^\prime}m_{
\tilde\ell^\prime}\right).\eqno(2.18)$$
Here, $m_{\tilde\ell}$ are integer variables defined on the links of the dual
lattice, which obey the constraint $\Delta_\mu^-m_\mu(\tilde x)=0$ (since
otherwise (2.18) would not be invariant under gauge transformations (2.15) on
$\phi$). These two facts constrain the $m$ to form closed
loops on the dual lattice, which interact via the Coulomb potential
as in NCQED. Once again the partition function may be thought of as a
sum over polymer configurations, though this time the polymers need be
neither singly charged nor self-avoiding. The
coupling strength is $g\equiv2\pi/e$; the excitations thus resemble
magnetic current loops with
charge specified by the quantisation condition (1.3).

We can confirm this picture of interacting magnetic current loops,
by coupling gauge-covariant fermions to the Villain model,
to yield a model we will call ``periodic QED'' (PQED)
to distinguish it from NCQED.
{}From what we already know, we can write
$$\eqalign{
Z_{PQED}=&\int{\cal D}\theta \det(D{\!\!\!\! /}\,[\theta]+\mu)
\sum_{\{n\}}\exp\left({{i\over2}n_{\mu\nu}(x)\Theta_{\mu\nu}(x)}
-{e^2\over4}n_{\mu\nu}^2(x)
\right)\cr
=&\sum_{\{{\cal
C}\}}\kappa^{-N_m}(-1)^{N_\Gamma}\left(\prod_{\ell\in\{\Gamma\}}\eta_\ell\right)
Z_{Villain}[\Gamma].\cr}\eqno(2.19)$$
Here, $Z_{Villain}[\Gamma]$ is the same partition function as that of
(2.12), but this time in the presence of an electric current distribution
defined by $\{\Gamma\}$. The steps (2.12-18) can be repeated to yield
$$\eqalign{
Z_{PQED}=&Z_{photon}[0]
\sum_{\{{\cal
C}\}}\kappa^{-N_m}(-1)^{N_\Gamma}\left(\prod_{\ell\in\{\Gamma\}}\eta_\ell\right)
\sum_{\{j\}}\delta_{j,\ell\in\{\Gamma\}}\sum_{\{S\}}
\delta_{j_\mu,\Delta_\nu^- S_{\nu\mu}}\cr
&\times\sum_{\{m\}}\exp\left(-{e^2\over2}j_\mu(x)v_{\mu\nu}(x-y)j_\nu(y)\right)
\exp\left(-{g^2\over2}m_\mu(\tilde x)v_{\mu\nu}(\tilde x-\tilde y)
m_\nu(\tilde y)\right)\cr
&\times\exp\left(-2\pi im_\mu(\tilde x)v_{\mu\nu}(\tilde x-\tilde y)
{1\over2}\epsilon_{\nu\lambda\rho\sigma}\Delta_\lambda^+S_{\rho\sigma}
(y+\hat\nu)\right).\cr}\eqno(2.20)$$
To implement the steps (2.12-18) it has been necessary to
introduce the oriented characteristic function
$S_{\mu\nu}$ for a surface which spans the self-avoiding loops in the
polymer expansion, but is otherwise unconstrained. Due to the
Kronecker $\delta$'s both $j$ and $S$ may be regarded as auxiliary
fields: a given configuration is uniquely specified by
${\cal C}$ and $m$\footnote{$^\dagger$}{This is not true in a finite volume --
we are grateful to John Stack for pointing this out}.
However, in the form (2.20) the similarity
between the electric and magnetic currents is suggestive. The model has a
new term -- the last exponential in (2.20) -- which describes the
interaction between electric and magnetic currents (the spacetime argument
of $S$ is determined by the requirement that the Coulomb propagator
couples to a conserved current at either end). To see that this is indeed
an interaction between magnetic and electric current distributions,
consider the distribution of a static magnetic charge $m_0(\tilde x)$ with
a spacelike surface $S_{ij}(y+\hat0)$, with $i,j$ ranging from 1 to 3.
The interaction may be written in the form
$igm_0(\tilde x)V_{mag}(\tilde x)$
where $V_{mag}$ is the {\it magnetic
scalar potential\/}:
$$V_{mag}(\tilde x)=\sum_y\left[-e{1\over2}\epsilon_{ijk}\Delta^+_i
v_{\rm 3d}(\tilde x-
\tilde y)S_{jk}(y+\hat0)\right] .\eqno(2.21)$$
The factor of $i$ is a consequence of the formulation in Euclidean space.
In the long wavelength limit the contribution to $V_{mag}$ from a current loop
spanned by a surface $S$ assumes its textbook form:
$$V_{mag}(x,S)={e\over{4\pi}}\int_{S}
{{\bf r.dS}\over r^3}=e{\Omega\over{4\pi}},\eqno(2.22)$$
where $\Omega$ is the solid angle subtended by the circuit at $x$.
Hence the interaction only depends on the boundary of the loop
$\partial S$. The magnetic
scalar potential is not single valued, since $\Omega$ is only defined modulo
$4\pi$: however, this has no physical consequence in the present case due to
the
Dirac quantisation condition, as we shall now discuss.

To see explicitly that the interaction only depends on the edge of the surface,
consider the following sequence of transformations, called the
``disentangling theorem'' in ref. [8],  on
the monopole-electric current interaction term in
Feynman gauge; we define $D_{\mu\nu}$ to be the integer-valued
characteristic function
for a Dirac sheet which spans the $m$ loop as $S$ spans the $j$ loop:
$$\eqalign{
&m_\mu(\tilde x)v_{\mu\nu}(\tilde x-\tilde
y){1\over2}\epsilon_{\nu\lambda\rho\sigma}\Delta_\lambda^+S_{\rho\sigma}
(y+\hat\nu)\cr
&=\Delta_\tau^-D_{\tau\mu}(\tilde x)v(\tilde x-\tilde y)
{1\over2}\epsilon_{\mu\lambda\rho\sigma}\Delta_\lambda^+S_{\rho\sigma}
(y+\hat\mu)\cr
&=-{1\over2}D_{\tau\mu}(\tilde x)\Delta_\tau^-\Delta_\lambda^+
v(\tilde x-\tilde y+\hat\tau+\hat\mu)\epsilon_{\mu\lambda\rho\sigma}
S_{\rho\sigma}(y)\cr
&=
{1\over2}D_{\tau\mu}(\tilde x)\Delta_\rho^-\Delta_\lambda^+
v(\tilde x-\tilde y+\hat\tau+\hat\mu)\epsilon_{\tau\mu\sigma\lambda}
S_{\rho\sigma}(y)
-{1\over4}D_{\tau\mu}(\tilde x)
\epsilon_{\tau\mu\rho\sigma}S_{\rho\sigma}(y)
\delta(\tilde x-\tilde
y+\hat\tau+\hat\mu)\cr
&=
-{1\over2}\Delta_\lambda^+D_{\tau\mu}(\tilde
x-\hat\lambda-\hat\mu-\hat\tau)v(x-y)\epsilon_{\tau\mu\sigma\lambda}
\Delta_\rho^-S_{\rho\sigma}(y)
-{1\over4}D_{\tau\mu}(\tilde x)
\epsilon_{\tau\mu\rho\sigma}S_{\rho\sigma}(x+\hat\tau+\hat\mu)\cr
&=
-j_\sigma(y)v(y-x){1\over2}\epsilon_{\sigma\lambda\tau\mu}\Delta_\lambda^+
D_{\tau\mu}(\tilde x-\hat\lambda-\hat\mu-\hat\tau)
-{1\over4}D_{\tau\mu}(\tilde x)\epsilon_{\tau\mu\rho\sigma}S_{\rho\sigma}
(x+\hat\tau+\hat\mu)
.\cr}\eqno(2.23)$$
We have used translation invariance $v(\tilde x)=v(x)=v(-x)$, and the identity
$$[\epsilon_{\lambda\tau\rho\sigma}\Delta_\mu^-\Delta_\lambda^++
   \epsilon_{\mu\lambda\rho\sigma}\Delta_\tau^-\Delta_\lambda^++
   \epsilon_{\mu\tau\lambda\sigma}\Delta_\rho^-\Delta_\lambda^++
   \epsilon_{\mu\tau\rho\lambda}\Delta_\sigma^-\Delta_\lambda^+]v(x)=
   -\epsilon_{\mu\tau\rho\sigma}\delta(x).\eqno(2.24)$$
We have also made the lattice coordinates explicit, resulting in a slightly
unwieldy expression. This is due to the offset of the dual lattice origin from
the direct one. More compact notations, which make use of the
language of differential forms, are available (eg. [7,22]).

The first thing to notice is that the expression ${1\over4}D\epsilon S$
is an integer (since $D$ and $S$ are themselves integer) -- the
intersection number of the two surfaces multiplied by the monopole and
electron charges. Since the coefficient of the term is $-2\pi i\equiv
-ieg$, this term contributes unity to $Z_{PQED}$ and hence has no
dynamical influence.
Therefore the electron-monopole interaction is
manifestly independent  of the {\it interior\/} of either of
the surfaces $S$ or
$D$ -- which are unconstrained by the dynamics -- but only their
boundaries, respectively the electric current $j$ or the magnetic
current $m$. It is important to note that these currents are constrained
to lie on direct and dual lattices, and that the surface independence,
which is a necessary requirement for the recovery of a local limit for
the quantum field theory, depends on the Dirac quantisation condition
(1.3). Essentially the same argument was put forward originally in [6].

The interaction betwen electric and magnetic charges is best described
on the lattice by a non-local interaction between conserved currents.
However, it is useful at this stage to pause and review the connection
between the formulation appearing in (2.20) and the more familiar description
in terms of local potentials and field strengths. {}From the discussion
preceding (2.21), we recall that magnetic charge has an interaction
energy depending on a magnetic scalar potential $V_{mag}$. By treating
$V_{mag}$ as the zeroth component of a four vector $\tilde A_\mu$,
it is possible to write the interaction term for electric current
as $ej_\mu A_\mu$, and the interaction for magnetic current as
$gm_\mu\tilde A_\mu$,
where the two potential fields $A_\mu$ and $\tilde A_\mu$ are given,
in continuum notation and in Feynman gauge, by
$$\eqalign{
A_\mu&=e{{-1}\over\partial^2}j_\mu+ig{1\over2}\epsilon_{\mu\lambda\rho\sigma}
\partial_\lambda{{-1}\over\partial^2}D_{\rho\sigma};\cr
\tilde A_\mu&=g{{-1}\over\partial^2}m_\mu-ie{1\over2}
\epsilon_{\mu\lambda\rho\sigma}\partial_\lambda{{-1}\over
\partial^2}S_{\rho\sigma}.\cr}\eqno(2.25)$$
Once again, the factors of $i$ associated with the $\epsilon$ symbols are
due to the Euclidean space formulation.
Field strength tensors $F_{\mu\nu}$ and $\tilde F_{\mu\nu}$ can now be defined
in the usual way by taking the four-dimensional curl of the potentials;
using (2.24) we find the relation between them:
$$-{{i}\over2}\epsilon_{\mu\nu\lambda\kappa}
(F_{\lambda\kappa}+eS_{\lambda\kappa})
=\tilde F_{\mu\nu}-gD_{\mu\nu}.\eqno(2.26)$$
Hence both $F$ and $\tilde F$ contain ambiguities due to Dirac sheets,
which in the continuum theory
must be resolved by the introduction of compensation terms supported
only on the sheets [2].
Of course, both surfaces $S$ and $D$ may be moved around at will by gauge
transformations, and hence have no physical meaning.
In continuum formulations of QEMD, in order to avoid double-counting, it
is customary to express the ambiguity solely in terms of $D_{\mu\nu}$,
but relations (2.25-26) expose the duality symmetry to the full.
On the lattice problems never arise so long as an action
which is a periodic function of the field strength, such as (2.12), is chosen;
the ambiguity then simply relates physically equivalent configurations
of the $F$ field (for the case of the Wilson action for U(1) lattice
gauge theory see [20]).

The final step in the construction of QEMD is to endow the monopole
excitations with specific dynamical  properties. In eqn. (2.20) the only
difference between monopoles and electrons are the constraints imposed
on the electric current loops by the polymer representation of the
fermion determinant, ie. that ${\cal C}$ be a partition of the lattice
into monomers, dimers and self-avoiding (and thus singly-charged)
polymers. Suppose we now insist that the dual lattice be similarly
partitioned by specifying $\tilde{\cal C}$ consisting of dual monomers,
dimers and polymers, which define loops of magnetic current. We then
write
$$\eqalign{
Z_{QEMD}=&Z_{photon}[0]\sum_{\{{\cal C}\}}\sum_{\{\tilde{\cal C}\}}
\kappa^{-N_m-N_{\tilde m}}(-1)^{N_\Gamma+N_{\tilde\Gamma}}
\left(\prod_{\ell\in\{\Gamma\}}\eta_\ell\right)
\left(\prod_{\tilde\ell\in\{\tilde\Gamma\}}\eta_{\tilde\ell}\right)\cr
&\times\sum_{\{j\}}\delta_{j,\ell\in\{\Gamma\}}\sum_{\{S\}}
\delta_{j_\mu,\Delta_\nu^-S_{\nu\mu}}
\sum_{\{m\}}\delta_{m,\tilde\ell\in\{\tilde\Gamma\}}\sum_{\{D\}}
\delta_{m_\mu,\Delta_\nu^-D_{\nu\mu}}\cr
&\times\exp\left(-{e^2\over2}j_\mu(x)v_{\mu\nu}(x-y)j_\nu(y)\right)
\exp\left(-{g^2\over2}m_\mu(\tilde x)v_{\mu\nu}(\tilde x-\tilde y)
m_\nu(\tilde y)\right)\cr
&\times\exp\left(2\pi im_\mu(\tilde x)v_{\mu\nu}(\tilde x-\tilde y)
{1\over2}\epsilon_{\nu\lambda\rho\sigma}\Delta_\lambda^+S_{\rho\sigma}
(y+\hat\nu)\right).\cr}\eqno(2.27)$$
Each configuration is completely specified by ${\cal C},\tilde{\cal C}$; the
variables $j$, $m$, $S$ and $D$ are auxiliary. Note that
the monopole fields have been given the same hopping parameter
$\kappa$ as the original fermion fields. Now, the crucial point is that
the model defined by (2.27) is invariant under the following duality
transformation:
$$j_\mu\leftrightarrow m_\mu\;\;\;\;\;\;\;\;e\leftrightarrow g\;\;\;\;\;\;\;\;
D\leftrightarrow
S\;\;\;\;\;
\;\;\;Z\leftrightarrow Z^*\eqno(2.28)$$
The manipulations (2.23) play a crucial role in demonstrating this. Note
further that under the interchange $m_\mu\leftrightarrow-m_\mu$, every term in
$Z_{QEMD}$ remains unchanged (since the number of links in every polymer
is even, the factors $\prod_{\tilde\ell}\eta_{\tilde\ell}$ remain
unaltered) except the argument of the last exponential, which changes
sign. Since this last term is a phase, and every $m$-polymer loop appears with
equal weight going in either direction, we find that
$$Z_{QEMD}\equiv Z^*_{QEMD}.\eqno(2.29)$$
Therefore the model is self-dual, respecting the symmetry (1.2):
in particular at the self-dual point
$e=g=\sqrt{2\pi}$ the electron and monopole degrees of freedom
behave identically.

To complete the formal exposition of the model, three final comments are
needed.
Firstly, although the role of the gauge potential has been suppressed in our
treatment, notice that the interactions in the model (2.27) are all described
in terms of conserved integer-valued currents; hence the model is clearly also
gauge invariant.
Next, although in the forms (2.6), (2.20) and (2.27) the various
partition functions $Z_{NCQED}$, $Z_{PQED}$ and $Z_{QEMD}$ consist of terms
with fluctuating sign, in the first two cases of NCQED and PQED we know
that the partition functions can also be expressed via
a positive
definite (albeit non-local) effective action, and hence are themselves positive
definite on a finite lattice. No such argument can be used for QEMD,
since it is not expressible simultaneously in terms of both local electron
and monopole fields. A demonstration of the positivity of $Z_{QEMD}$ remains
an open question, and is desirable for the discussion of
the possible phase structure
of the model (see next section).
Thirdly, in the form (2.27) the magnetic current $m_\mu$ transforms as a
vector, by analogy with the electric current.
Even in the classical theory descibed by (1.1), it is impossible
to satisfy invariance simultaneously under discrete parity (P) and
charge-conjugation (C) transformations.
For a vector $m_\mu$, under P:
$$\eqalign{
{\bf E}\mapsto-{\bf E}\;\;\;&;\;\;\;{\bf B}\mapsto{\bf B};\cr
(\rho_e,{\bf j}_e)\mapsto(\rho_e,-{\bf j}_e)\;\;\;&;
(\rho_m,{\bf j}_m)\mapsto(\rho_m,-{\bf j}_m),\cr}\eqno(2.30)$$
whereas under C:
$$\eqalign{
{\bf E}\mapsto-{\bf E}\;\;\;&;\;\;\;{\bf B}\mapsto-{\bf B};\cr
(\rho_e,{\bf j}_e)\mapsto(-\rho_e,-{\bf j}_e)\;\;\;&;\;\;\;
(\rho_m,{\bf j}_m)\mapsto(-\rho_m,-{\bf j}_m).\cr}\eqno(2.31)$$
The QEMD Maxwell equations (1.1) are invariant under C but not P, or hence CP.
If $m_\mu$ is defined to be an axial vector, then under P:
$$(\rho_m,{\bf j}_m)\mapsto(-\rho_m,{\bf j}_m),\eqno(2.32)$$
and under C:
$$(\rho_m,{\bf j}_m)\mapsto(\rho_m,{\bf j}_m),\eqno(2.33)$$
in which case (1.1) are now invariant under P, but not C or CP. A
Euclidean lattice
analogue of this latter case can also be constructed if the monopole
conserved current is derived using a  kinetic term of the form
$$\eqalign{
(D{\!\!\!\! /}\,+\mu)(x,y)&={1\over2}\sum_\mu\eta_\mu(x)\varepsilon(x)\left[
\delta_{y,x+\hat\mu}\exp(ie\theta_\mu(x))
-\delta_{y,x-\hat\mu}\exp(-ie\theta_\mu(x-\hat\mu)\right]\cr
&+\mu\delta_{y,x}\varepsilon(x).\cr}\eqno(2.34)$$
Here, $\varepsilon(x)$ is the alternating phase $(-1)^{x_1+x_2+x_3+x_4}$.
The result of this modification in the polymer expansion is to reverse
the sign of every loop $\Gamma$ containing $(2n+2)$ links.
\vskip 1.0 truecm
\noindent{\bf3. Phase Structure}

In this section we will try to discuss the lattice
model described in section 2 more quantitatively,
with particular emphasis on its phase structure.
We shall draw insight and information from what is already known about
NCQED, Villain QED, and PQED. First, however, we shall consider the limit
of large fermion bare mass $\mu$, in which case an expansion in powers
of $1/\mu$ is viable [23].

For large $\mu$, the partition function in (2.27) is dominated by
configurations
in which $N_m$ and $N_{\tilde m}$, the numbers of direct and dual monomers
respectively, are large. Both dimers and polymers must be considered as
excitations suppressed by powers of $\kappa=1/2\mu$. Only polymers, however,
interact with the electromagnetic field, and hence with other loops,
via the exponential terms of (2.27). Consider the smallest monopole loop
excitation, simply four dual links bounding a dual plaquette. By considering
the photon exchange diagrams of figure 1, and using the Feyman gauge
expression
$$v_{\mu\nu}(\tilde x-\tilde y)=\delta_{\mu\nu}\int_p
{{\exp(ip.(x-y))}\over S(p)},\eqno(3.1)$$
with
$$\int_p\equiv\int_{-\pi}^\pi{{d^4p}\over{(2\pi)^4}}$$
and
$$S(p)=4\sum_\mu\sin^2\left({1\over2}p_\mu\right),$$
we obtain the result that a single loop excitation is suppressed by the
factor $\kappa^4\exp(-g^2/4)$ [23]. We have folded in the other factors from
(2.27): -1 for a fermion loop, -1 from the signed KS phase, and $\kappa^4$
from the hopping parameters. Now, consider the effect of such a loop on a
photon propagating between electric currents on the direct lattice at sites
0 and $y$. {}From the disentangling transformation (2.23) we know that electric
current interacts with the four-dimensional curl of the Dirac sheet, which for
the simple dual plaquette excitation is a plaquette on the direct lattice.
We can then consider the process shown in figure 2.

Summing over all possible positions and orientations of the loop, we find for
the contribution $v_{\mu\nu}^{[1]}$ to the ``dressed'' propagator:
$$\eqalign{
v_{\mu\nu}^{[1]}(0,y)=2g^2\sum_x\sum_{\alpha>\beta}v_{\mu\alpha}(0,x)\times
&[2v_{\alpha\nu}(x,y)-v_{\alpha\nu}(x+\hat\beta,y)
-v_{\alpha\nu}(x-\hat\beta,y)-v_{\beta\nu}(x,y)\cr
&+v_{\beta\nu}(x+\hat\alpha,y)
+v_{\beta\nu}(x-\hat\beta,y)-v_{\beta\nu}(x+\hat\alpha-\hat\beta,y)].\cr}
\eqno(3.2)$$
It is most convenient to evaluate (3.2) in Landau gauge;
$$v_{\mu\nu}(x,y)=\int_p{{\exp(ip.(x-y))}\over S(p)}{\cal P}_{\mu\nu}(p),
\eqno(3.3)$$
with the lattice transverse projection operator ${\cal P}_{\mu\nu}(p)$ given
by
$${\cal P}_{\mu\nu}(p)=\delta_{\mu\nu}-{{\sin\left({1\over2}p_\mu\right)
\sin\left({1\over2}p_\nu\right)}\over
{\sum_\rho\sin^2\left({1\over2}p_\rho\right)}}.\eqno(3.4)$$
We find
$$v_{\mu\nu}^{[1]}(0,y)=g^2\int_p{{e^{-ip.y}}\over S(p)}
\left\{{\cal P}_{\mu\nu}(p)-\sum_{\alpha\beta}{\cal P}_{\mu\alpha}(p)
{\cal P}_{\beta\nu}(p){{(1-e^{-ip_\alpha})(1-e^{ip_\beta})}\over S(p)}\right\}.
\eqno(3.5)$$
In the long wavelength limit $p\to0$ the second term in braces is $O(p^2)$
and hence irrelevant. Combining this result with the suppression factor
of figure 1, we obtain for the full photon propagator
$v_{\mu\nu}=v_{\mu\nu}^{[0]}+v_{\mu\nu}^{[1]}$:
$$\lim_{y\to\infty}v_{\mu\nu}(0,y)=\int_p{{e^{-ip.y}}\over p^2}
{\cal P}_{\mu\nu}(p)\left(1+g^2\kappa^4\exp\left(-{g^2\over4}\right)\right).
\eqno(3.6)$$

Expression (3.6) is of the same form as the bare continuum propagator, except
that the overall strength of the interaction has been rescaled, that is,
electric charge has been renormalised: $e^2_R=Z_ge^2$, with the renormalisation
constant $Z_g$ given by
$$Z_g=1+{g^2\over{8\mu^4}}\exp\left(-{g^2\over4}\right).\eqno(3.7)$$
This expression is the first term of an expansion in $1/\mu$, but is
non-perturbative (and exact) in $g$ (it is also worth noting that since
the fermion determinant is polynomial in $\mu$, the expansion in
$1/\mu$ is finite, and there are no non-perturbative contributions).
The most interesting aspect is the $+$
sign of the correction, which is in contrast to the charge renormalisation
due to an electric current loop found originally for NCQED in [23]:
$$Z_e=1-{e^2\over{8\mu^4}}\exp\left(-{e^2\over4}\right).\eqno(3.8)$$
Of course, the difference has its origin in the factor $ieg$ in the
interaction between electric and magnetic currents, and supports the claim
made in the introduction that monopoles anti-screen electric charge.
Indeed, at the self-dual point $e=g$, $Z_e=Z_g^{-1}$ and bare charges remain
unrenormalised to this order. This agrees
with the results of Calucci and
Jengo [7], who considered an expansion not in inverse mass but in number
of fluctuating loops. Because they considered scalar matter fields, they
were able to use a proper time formalism and consider the
current loops as continuum worldlines. They also found that a charge
renormalisation was
caused by the effect of small intermediate loops (ie. from the
limit $T\to0$, where $T$ parametrises the length of the worldline loop),
although the precise details
of their regularisation are left hazy. It would be interesting to pursue the
inverse mass expansion further to see if the self-duality persists (indeed,
the expansion was originally
suggested as a means of treating fermion fields in [7]).
For larger magnetic loop excitations, the dual loop interacting with
the electric current is no longer the same shape; indeed, a large planar
monopole loop corresponds to an array of elementary electric current
plaquettes in the orthogonal directions.
It seems highly plausible, however,  that
Calucci and Jengo's geometrical arguments
that electric
and magnetic loops of the same size and shape will induce opposite charge
renormalisations will continue to hold in the limit that loop size is much less
than photon wavelength, although the fluctuating signs due to the fermionic
nature of the currents may present a complication.

Now, in order to define a continuum field theory, we need to approach
the continuum limit, in which all particles are light; the $1/\mu$ expansion
cannot help in this regard. Next we consider the limit $e\to0$, $g\to\infty$.
By self-duality, our findings will also hold in the dual limit
$e\to\infty$, $g\to0$. For $g$ sufficiently large (see below), the
exponential term
$\exp(-{g^2\over2}m_{\tilde\ell}v_{\tilde\ell\tilde\ell^\prime}
m_{\tilde\ell^\prime})$ in (2.27) alone will suppress monopole polymers,
and the monopole sector will be described entirely in terms of monomers
and dimers, which do not interact with the electromagnetic field. In this
limit the argument of the third exponential term in (2.27) describing
the monopole-electron interaction also vanishes; only loops which subtend
a non-zero solid angle feel this term. Therefore $Z_{QEMD}$ factorises:
$$\lim_{e\to0,g\to\infty}Z_{QEMD}=Z_{NCQED}(e)Z_{NCQMD}(g).\eqno(3.9)$$
By reversing the sequence of transformations leading to eqn. (2.11) we
see that the electron sector is now governed solely by NCQED (2.1), which
may be adequately treated in perturbation theory. The staggered lattice
fermion formulation used here is known to describe four physical fermion
species; in the long wavelength limit we recover continuum QED with
$N_f=4$.
The monopole sector, on
the other hand, is saturated by monomers and dimers.
It is known rigorously [24] that in the chiral limit $\mu\to0$
($\kappa\to\infty$) the chiral symmetry of the monopole degrees of
freedom is spontaneously broken. In continuum field theory, this is
signalled by a non-vanishing condensate $\langle\bar\psi\psi\rangle$,
where $\psi,\bar\psi$ are monopole field operators. In the polymer
language this symmetry-breaking order parameter is given by [21,24]
$$\displaystyle\lim_{\mu\to0}{\langle\bar\psi\psi(\mu)\rangle}=
{1\over V}{{\partial\ln Z_{NCQMD}}\over{\partial \mu}}\biggr\vert_{\mu=0}=
\displaystyle\lim_{\mu\to0}{1\over V}{{\langle N_{\tilde m}\rangle\over \mu}}.
\eqno(3.10)$$
Symmetry breaking is signalled by the equilibrium concentration of
monomers vanishing with some power of $\mu$ less than or equal to one.

In the staggered fermion formulation, global chiral transformations
are defined by a U(1) rotation:
$$\psi(x)\mapsto\exp(i\alpha\varepsilon(x))\psi(x)\;\;\;;\;\;\;
\bar\psi(x)\mapsto\exp(i\alpha\varepsilon(x))\bar\psi(x).\eqno(3.11)$$
Hence in the polymer
representation these rotations leave both polymers and dimers uneffected,
since each contain an equal number of odd and even sites.
Electron and monopole fields can be
independently rotated -- so in the $\mu\to0$ limit $Z_{QEMD}$ has a
${\rm U(1)}_V\otimes{\rm U(1)}_A\otimes{\rm U(1)}_{\tilde V}\otimes{\rm
U(1)}_{\tilde A}$ symmetry, with $V$ and $A$ standing for vector and axial
vector symmetries respectively on both direct and dual lattices.
Thus for $e\to0$ $g\to\infty$, the condensate $\langle N_{\tilde m}\rangle$
breaks the symmetry to
${\rm U(1)}_V\otimes{\rm U(1)}_A\otimes{\rm U(1)}_{\tilde V}$.
Of course, by Goldstone's theorem there must in this case be a massless
boson in the spectrum -- this will be a tightly bound monopole --
anti-monopole state which we shall refer to as a ``magnetopion''. In the
$g\to\infty$, $\mu\to0$
limit in which the monopole physics is described by dimer
configurations, the magnetopions have vanishing spatial extent, and
do not couple either to photons or electrons.

For $g$ large but finite, one may once again estimate the effect of
small monopole loop excitations on charge renormalisation, using a
mean field argument. If the vacuum consists of a fraction $\rho$
of monomers ($\rho=1$ in the $1/\mu$ expansion), then the excitation cost
of a monopole loop changes to $(2\mu)^{4\rho}$. Using (3.10), the expression
for $Z_g$ becomes
$$Z_g=1+2g^2\exp\left(-{g^2\over4}+4\langle\bar\psi\psi\rangle\mu\ln{1\over\mu}
\right).\eqno(3.12)$$
Note that the correction involving $\langle\bar\psi\psi\rangle$ vanishes
in the chiral limit.
For electric current loops we can use the perturbative QED result
$$Z_e=1-{e^2\over{6\pi^2}}\ln{1\over\mu}.\eqno(3.13)$$
Since  the loop suppression is now due to the factor $\exp(-g^2/4)$, eqn.
(3.12) is no longer the first term in a systematic expansion; the next
smallest loop is suppressed by a factor $\exp(-0.4311g^2)$ [23].

Before turning to the behaviour of QEMD for intermediate values of
$e$, $g$, it will be useful to
discuss and contrast the phase structure of the two other models we have
introduced, NCQED (2.1), and Villain QED (2.12).
As mentioned above, in the long wavelength limit NCQED resembles
continuum QED with $N_f=4$. Numerical simulations of this model reveal
that in the limit $\mu\to0$ it exhibits a continuous chiral
symmetry breaking phase transition for $e$ in the range 2.2 - 2.3 [13,14].
The Villain model, which has no fermion fields, has a phase transition,
possibly weakly first order,
at $e\simeq1.25$ [25], or $g\simeq5.0$, separating confinement and
Coulomb phases. There is no local order
parameter due to Elitzur's theorem.

At first sight, it is not clear in this language why either model has a
phase transition. For large $x$, $v(x)\propto1/x^2$, so that any change
in the coupling strength in (2.9) or (2.18) can simply be absorbed by a change
of length scale. Thus for weak
coupling, Villain QED resembles continuum QED, and electric charges
interact via Coulomb's law.
However, for small $x$, the lattice cutoff breaks this
scaling symmetry; eg. $v(0)$ is actually a finite quantity. This means
that in the Villain model, once $g$ is sufficiently small the monopole
excitations of like sign no longer have a strong repulsive interaction,
and those of unlike sign are no longer strongly attracted. At this point
hitherto small monopole loops become free to grow and spread over large
distances, forming a plasma which screens the long-range Coulomb
forces. If a Wilson loop of electric current is introduced into the
plasma, the interaction term $\exp(-i2\pi m_{\tilde\ell}v_{\tilde\ell
\tilde\ell^\prime}(\epsilon\Delta S)_{\tilde\ell^\prime})$ has the
effect of disordering its phase, and causing its expectation value to
decay as its area, signalling confinement. This is a non-perturbative
generalisation of the anti-screening effect discussed above.
For a loop of length $L$ the activation energy $\sim g^2Lv(0)$, and the entropy
$\sim L\ln7$ -- so the critical $g$ can be
estimated by an energy/entropy argument [19]. This ``monopole
condensation'' is now widely accepted as being the mechanism behind the
phase transition in not only the Villain model but also U(1) lattice
gauge theory with the Wilson action -- there are similarities with the
Kosterlitz-Thouless-Berezinsky mechanism in the two dimensional X-Y
model.

Now consider NCQED (2.9). This model has the same driving term describing a
Coulomb gas of current loops, but with three important modifications.
First, the loops are constrained to be singly-charged and self-avoiding.
Secondly, there is the sign factor. Thirdly, even for a given loop
configuration $\{\Gamma\}$ there will be many different background
monomer/dimer configurations, each carrying its own weight
$\kappa^{-N_m}$. In the limit $e\to\infty$ all loops are suppressed, and
we are left with a pure monomer/dimer system which is known to break
chiral symmetry [24]. In the limit $e\to0$ we are left with free
massless fermions -- hence the correlation length must diverge. In this
limit large loops are unsuppressed, although it is far from clear
whether they must dominate $Z_{NCQED}$ -- indeed since the contributions
to $Z$ contain fluctuating signs an entropy argument cannot be constructed.
Thus, it is unclear whether the phase transition in this model
coincides with loops growing without bound -- so much is hidden in the
sign factor. Even continuum QED
can potentially exhibit singular behaviour, via the unbounded
$\sigma_{\mu\nu}F_{\mu\nu}$ term which appears in the expression for
the inverse propagator [26].

Now we return to QEMD;
for intermediate values of
$e$, $g$, the interaction between electrons and monopoles governed
by the final exponential in (2.27) becomes crucial. As discussed above,
the standard lore is
that monopole loops disrupt the phases of electron current loops,
causing contributions from larger electric loops to cancel in the
partition function, and hence suppressing these loops, leading to
electric charge confinement. Similarly, electron loops will
disrupt magnetic current loop phases. We may then question the existence
of a phase in which large polymers are present simultaneously on both
direct and dual lattices (which would correspond to the coexistence of
light electrons and monopoles). Even though the loops each carry
a sign from the polymer expansion of the determinant, making each term in the
partition function of indefinite sign, the additional random phase due to
the interaction must surely effectively cancel all contributions to
$Z_{QEMD}$ of this form.
This physical interpretation is based on the idea that the
polymers are the nearest we can get to a ``fermion worldline''.
We should also not forget the technical obstacle,
discussed in section 2,
namely a
demonstration that $Z_{QEMD}$ is positive for real values of $e$.

To summarise our conclusions so far: there is probably no extended region
of the phase diagram where both electrons and magnetic monopoles are light
propagating particles.
For $e\to0$ the only light degrees
of freedom are electrons and decoupled magnetopions,
and vice versa for $g\to0$.
For $g$ large but finite, the monopoles' chiral condensate
$\langle\bar\psi\psi\rangle$ is non-vanishing in lattice units, and
hence divergent in physical units. By analogy with hadronic physics,
therefore, we expect the decay constant $f_\pi$ to diverge and the
pion screening length $m_\pi/f_\pi^2$ to vanish in the chiral limit,
leaving a theory of non-interacting bosons in the monopole sector.
The question of the magnetopions'
coupling to photons and hence electrons is more delicate, depending on
the spatial extent of the polymer configurations dominating $Z_{QEMD}$,
but from the previous discussion there is no particular reason to suppose
that the coupling will remain non-vanishing in the continuum limit
if $\langle\bar\psi\psi\rangle>0$
The only scenario which supports
an interacting continuum limit for QEMD, therefore, is one in which the
chiral symmetry of the electrons is spontaneously broken
precisely at the self-dual
point $e=g=\sqrt{2\pi}$; by duality the monopoles' chiral symmetry will be
restored at precisely the same point. It may thus be
possible to approach the self-dual point from either phase to take
a continuum limit, in which the electrons are in their chirally
symmetric phase (say), with physical mass vanishing in the limit
$\mu\to0$, and the monopoles are in a chirally broken state,
with the condensate $\langle\bar\psi\psi\rangle$ vanishing non-analytically
as $e\to(\sqrt{2\pi})_-$, and some related mass scale defining an
inverse correlation length, giving dimensional transmutation
{\it \`a la\/} Miranskii [11].
Of course, the self-dual point is also a candidate
for a zero of the $\beta$-function using the arguments following (3.8).

Unfortunately, the attractive scenario of a fixed point at the self-dual
point appears to be excluded by current numerical data. We have already
noted that in the limit $g\to\infty$ the monopole loops decouple to
leave standard lattice NCQED. As discussed above, NCQED is known to exhibit a
chiral symmetry breaking phase transition at $e\simeq2.2-2.3$, which is
some way short of the self-dual value $e=2.5066\ldots$
It is difficult to believe (and indeed runs contrary to the arguments
presented above) that the effect of monopole - electron interactions
will revise this value upwards: if anything monopoles make the
$e^+e^-$ interaction stronger, and would thus {\it reduce\/}
the value of the
critical coupling needed for chiral symmetry breakdown.
Hence we must conclude that at the
self dual point electrons and monopoles probably both have their chiral
symmetry
spontaneously broken. The resulting model is necessarily massive,
the correlation length thus finite, and
the full $\beta$-function non-vanishing at this point. Instead
there must be a region of non-zero width centred on the self-dual
point, separating two phases where
light particles can exist at two critical couplings $e_c$, $2\pi/e_c$.

One final point should be made. As mentioned in the introduction, the
considerations of this paper were originally motivated by the observation
that monopole-like excitations, defined on the dual lattice, become
dense in a geometric sense and percolate very close to or actually
at the point where chiral symmetry is broken [13,18,27]. This observation
encouraged us to speculate that monopole condensation was the agent behind
chiral symmetry breaking in NCQED [26].
The truth cannot be so simple -- as pointed
out in [28], the percolation transition does not correspond to a
condensation in the Bose-Einstein sense (assuming bosonic monopoles), and
indeed the monopole excitations of NCQED do not have the long range
potentials that could have the effects discussed here. Only formulations
in which the action is periodic in the field strength, and hence in which
the Dirac string is invisible, can give rise to the monopole condensation
phenomenon usually discussed. Moreover the monopole-like excitations
of [18] will {\sl still\/}
be present as dislocations in the photon field of the NCQED
sector of the lattice QEMD model,
independent of the dynamical monopoles we have introduced
explicitly. Therefore it seems unlikely that the dynamical
monopoles described here
could be the primary agent of chiral symmetry breaking in QEMD, at least for
$N_f=4$. It is interesting to note that
the numerical coincidence of chiral and monopole percolation
transitions at higher values of $N_f$ remains impressive [29].
Moreover, the critical exponent $\nu$ derived from the size of the
largest cluster is consistent with the critical indices found in
powerlaw fits of the chiral equation of state
of NCQED [29]. Therefore
it may well be that the monopoles seen in NCQED are connected
in some way with chiral symmetry breaking.  Quenched
simulations using alternative actions for the lattice photon field
manage to separate the transitions, but still suggest
that chiral symmetry
breakdown is due to fluctuations on the scale of the lattice
spacing [16].
\vskip 1.0 truecm
\noindent{\bf4. Discussion}

We have succeeded in formulating QEMD using an explicit regularisation
which preserves an electromagnetic duality symmetry. This has been done
in effect by introducing an interaction term into the expansion of the
partition function, rather than directly into the action. Our formulation
(2.27) shares the non-local nature of previous approaches, but is more
economical in the sense that the Dirac quantisation condition necessary
to ensure covariance of Green functions is minimal, viz:
$$eg=2\pi.\eqno(4.1)$$
This follows from the natural appearance of monopole-like excitations
in abelian lattice gauge theory [19]. The lattice formulation admits
non-perturbative approaches such as the $1/\mu$ expansion discussed in
section 3. Although we have not pursued this in detail, it seems plausible
that higher order calculations will prove to be consistent with the
claim of Calucci and Jengo that monopole and electron virtual loops
give radiative corrections to the photon propagator of opposite sign, and
hence that there will be no charge renormalisation at the self-dual point
$e=g$.

On the negative side, the fermionic nature of the monopoles, which we
impose precisely in order to have a self-dual symmetry, means that it
remains to be proved whether $Z_{QEMD}(e)>0$ for $e$ real. Moreover,
the fluctuating signs in the terms of $Z_{QEMD}$ mean that Monte Carlo
simulations are probably impracticable; numerical studies of
low-dimensional fermionic theories
using the polymer representation, and even free fermions, have proved difficult
[21,30]. Despite this, one can argue that results from simulations of NCQED
suggest that QEMD in the form (2.27) does not have a continuum limit at the
self-dual point, and that there is no value of $e$ for which both electrons
and monopoles are light particles.

The reason for the difficulty of finding a continuum limit is probably
the excitation energies for polymer loops discussed in section 3. Even the
smallest loop, a single plaquette, is suppressed by a factor
$\exp(-e^2/4)\simeq0.21$ at the self-dual point -- large loops necessary for
the description of light particles are even further suppressed. The
suppression factors are intimately bound up in the geometry of the
lattice regularisation; it is difficult to see how the lattice could be
modified without also destroying the duality symmetry.

There are other possible modifications. Recall that the reason a continuum
limit appears elusive is that the critical coupling $e_c$ of NCQED is
somewhat smaller than $\sqrt{2\pi}$. Could NCQED be modified to yield a larger
$e_c$? One could introduce a repulsive four-fermi interaction to
delay the onset of spontaneous chiral symmetry breaking.
The suggested lattice interaction
is [31]
$$\eqalign{
S_{rep}&=\lambda\sum_{xy}\bar\chi(x)M(x,y)\chi(y),\cr
M(x,y)&=\sum_\mu\delta_{y,x+\hat\mu}\exp(i\phi_\mu(x))+
  \delta_{y,x-\hat\mu}\exp(-i\phi_\mu(x-\hat\mu)),\cr}\eqno(4.2)
$$
where $\chi,\bar\chi$ are generic fermion fields and $\phi_\mu$ is a vector
auxiliary field, defined on the lattice links, which is freely integrated
over. The $+$ sign in the definition of $M$ is responsible for the
repulsion; it renders $M$ hermitian rather than antihermitian, and hence
forces the eigenvalues of the full fermion kinetic operator to be complex.
Repulsive fermion -- anti-fermion interactions are not usually considered
in isolation due to Dyson's argument about vacuum stability [32] -- in the
context of a strongly-coupled theory these arguments are hopefully
inappropriate.

A second possibility to increase $e_c$ is to increase the number $N_f$
of fermion flavors in the model. Simulations with varying $N_f$ [29,33]
show clear evidence that $e_c$ increases with $N_f$. For sufficiently large
$N_f$ (estimated between 13 [33] and 24 [29]) the phase transition becomes
first order. However, it is conceivable that there is a window of
$8\leq N_f\leq15$ where both the transition is continuous and $e_c>\sqrt{2\pi}$
(ie. $1/e_c^2\equiv\beta_c^2=0.159\ldots$). In this case introduction of
a dual sector of magnetic monopoles could force the existence of a phase
transition, and hence a fixed point, at the self-dual point.

Of course, even if no interacting continuum limit can be found, the model
presented here may be of interest as an effective theory of strongly
interacting and massive fermionic
abelian monopoles arising from some more fundamental
model. It is possible in this case to relax the condition that
$\mu_{monopole}=\mu_{electron}$. It is both interesting and amusing to
contemplate the phenomenological implications of lattice QEMD. A recent
survey of experimental and observational limits on monopole matter, and
in particular the effect of virtual monopole loops, has been
given in [34]. The novelty of the present approach is that monopoles are
strongly interacting fermions, which means that in our world of small
$e\simeq0.3$, monopole chiral symmetry is spontaneously broken, and there
will be at least one light Goldstone particle in the monopole sector,
the magnetopion $\pi_m$. Can magnetopions interact with
ordinary matter? The scattering of photons off tiny magnetic dipole moments
will depend on the details of the magnetic charge distribution in the
magnetopion, which in turn depends on the (non-perturbative) dynamics.
However, we note that magnetopion production via virtual photon decay
$\gamma\to n\pi_m$ is forbidden by charge-conjugation symmetry, just as
strong interaction decays such as $\rho^0\to n\pi^0$ are forbidden. The
main interaction of magnetopions, just as for axions, will be via the
coupling $\pi_m\gamma\gamma$; indeed, massive magnetopions will decay to
two photons. Since the lattice regularisation presented here does not admit
an axial anomaly, magnetopion decay will be suppressed by a factor
of $O(m_{\pi_m}^2)$ by the usual arguments [35]. This raises the interesting
question of whether there is a range of bare mass $\mu$ such that the
magnetopion is both sufficiently long-lived and sufficiently heavy to be of
cosmological significance. Such issues cannot be resolved without a further
understanding of the detail of the strongly interacting monopole dynamics.
\vskip 1 truecm
\noindent{\bf Acknowledgements}

The work of SJH was supported partly by a CERN Fellowship and partly by a PPARC
Advanced Fellowship. JBK is supported in part by the National Science
Foundation,
NSF PHY92-00148.
\vfill\eject
\noindent{\bf References}

\noindent
[1] P.A.M. Dirac, Proc. R. Soc. {\bf A133} (1931) 60.

\noindent
[2] M. Blagojevi\'c and P. Senjanovi\'c, Phys. Rep. {\bf157} (1988) 233.

\noindent
[3] N. Cabibbo and E. Ferrari, Nuovo Cimento {\bf23} (1962) 1647.

\noindent
[4] J. Schwinger, Phys. Rev. {\bf144} (1966) 1087.

\noindent
[5] D. Zwanziger, Phys. Rev. {\bf D3} (1971) 880.

\noindent
[6] R.A. Brandt, F. Neri and D. Zwanziger, Phys. Rev. {\bf D19} (1979) 1153.

\noindent
[7] G. Calucci and R. Jengo, Nucl. Phys. {\bf B223} (1983) 501.

\noindent
[8] G. Calucci, R. Jengo and M.T. Vallon, Nucl. Phys. {\bf B211} (1983) 77.

\noindent
[9] J. Schwinger, Phys. Rev. {\bf151} (1966) 1048, 1055;\hfill\break
R.Brandt and F. Neri, Phys. Rev. {\bf D18} (1978) 2080;\hfill\break
C. Panagiotakopoulos, J. Phys. {\bf A16} (1983) 133.

\noindent
[10] L.D. Landau and I.Ya. Pomeranchuk, Dokl. Akad. Nauk. SSSR {\bf102}
(1955) 489;\hfill\break
L.D. Landau, in {\sl Niels Bohr and the Development of Physics\/} (ed.
W. Pauli) p.52, Pergamon Press, London (1955).

\noindent
[11] P.I. Fomin, V.P. Gusynin, V.A. Miranskii and Yu.A. Sitenko, Riv.
Nuovo Cimento {\bf6} (1983) 1;\hfill\break
V.A. Miranskii, Nuovo Cimento {\bf90A} (1985) 149.

\noindent
[12] J.B. Kogut, E. Dagotto and A. Koci\'c, Phys. Rev. Lett. {\bf60} (1988)
772; Nucl. Phys. {\bf B317} (1989) 271.

\noindent
[13] A. Koci\'c, J.B. Kogut and K.C. Wang, Nucl. Phys. {\bf B398} (1993) 405.

\noindent
[14] M. G\"ockeler, R. Horsley, P.E.L. Rakow, G. Schierholz and R. Sommer,
Nucl. Phys. {\bf B371} (1992) 713.

\noindent
[15] S.J. Hands, J.B. Kogut and J.H. Sloan, Nucl. Phys. {\bf B344} (1990)
255; Erratum, Nucl. Phys. {\bf B352} (1991) 528.

\noindent
[16] S.J. Hands, Nucl. Phys. {\bf B}(Proc. Suppl.){\bf42} (1995) 663.

\noindent
[17] C.N. Leung, S.T. Love and W.A. Bardeen, Nucl. Phys. {\bf B273} (1986)
649.

\noindent
[18] S.J. Hands and R.J. Wensley, Phys. Rev. Lett. {\bf63} (1989) 2169.

\noindent
[19] T. Banks, R. Myerson and J.B. Kogut, Nucl. Phys. {\bf B129} (1977)
493.

\noindent
[20] T.A. DeGrand and D. Toussaint, Phys. Rev. {\bf D22} (1980) 2478.

\noindent
[21] M. Karowski, R. Schrader and H.J. Thun, Commun. Math. Phys. {\bf97}
(1985) 5.

\noindent
[22] A.H. Guth, Phys. Rev. {\bf D21} (1980) 2291;\hfill\break
L. Polley and U.-J. Wiese, Nucl. Phys. {\bf B356} (1991) 629.

\noindent
[23] S.W. de Souza and R.D. Kenway, Phys. Lett. {\bf B248} (1990) 423;
Nucl. Phys. {\bf B354} (1991) 39.

\noindent
[24] P. Rossi and U. Wolff, Nucl. Phys. {\bf B248} (1984) 105;
\hfill\break
M. Salmhofer and E. Seiler, Commun. Math. Phys. {\bf 139} (1991)
395.

\noindent
[25] R.J. Wensley, Ph.D. thesis, University of Illinois (1989);\hfill\break
Z. Schram and M. Teper, Phys. Rev. {\bf D48} (1993) 2881.

\noindent
[26] S.J. Hands, J.B. Kogut and A. Koci\'c, Nucl. Phys. {\bf B357} (1991) 467.

\noindent
[27] S.J. Hands, A. Koci\'c, J.B. Kogut, R.L. Renken, D.K. Sinclair and
K.C. Wang, Phys. Lett. {\bf B261} (1991) 294; Nucl. Phys. {\bf B413} (1994)
503.

\noindent
[28] P.E.L. Rakow, Nucl. Phys. {\bf B}(Proc. Suppl.){\bf30} (1992) 591;\hfill
\break
M. G\"ockeler, R. Horsley, P.E.L. Rakow and G. Schierholz, preprint
DESY-93-025, hep-lat/9303001 (1993).

\noindent
[29] J.B. Kogut and K.C. Wang, Illinois preprint ILL-TH-95-\#28,
hep-lat/9501021 (1995).

\noindent
[30] I. Montvay, Phys. Lett. {\bf B227} (1989) 260; in  proceedings of
the 1989 Carg\`ese Workshop {\sl Probabilistic Methods in Quantum Field
Theory and Quantum Gravity\/}, p.87 (1989).

\noindent
[31] J.B. Kogut and J.-F. Laga\"e, Nucl. Phys. {\bf B}(Proc. Suppl.){\bf42}
(1995) 681.

\noindent
[32] F.J. Dyson, Phys. Rev. {\bf85} (1952) 631.

\noindent
[33] E. Dagotto, A. Koci\'c and J.B. Kogut, Phys. Lett. {\bf B232} (1989) 235;
\hfill\break
V. Azcoiti, G. Di Carlo and A.F. Grillo, Phys. Lett. {\bf B305} (1993) 275.

\noindent
[34] A. De R\'ujula, Nucl. Phys. {\bf B435} (1995) 257.

\noindent
[35] D.G. Sutherland, Nucl. Phys. {\bf B2} (1967) 433;\hfill\break
M. Veltman, Proc. R. Soc. {\bf A301} (1967) 107;\hfill\break
S.L. Adler, Phys. Rev. {\bf177} (1969) 2426.

\vfill\eject
\input psfig
\vfill
\vbox{
\centerline{
\psfig{figure=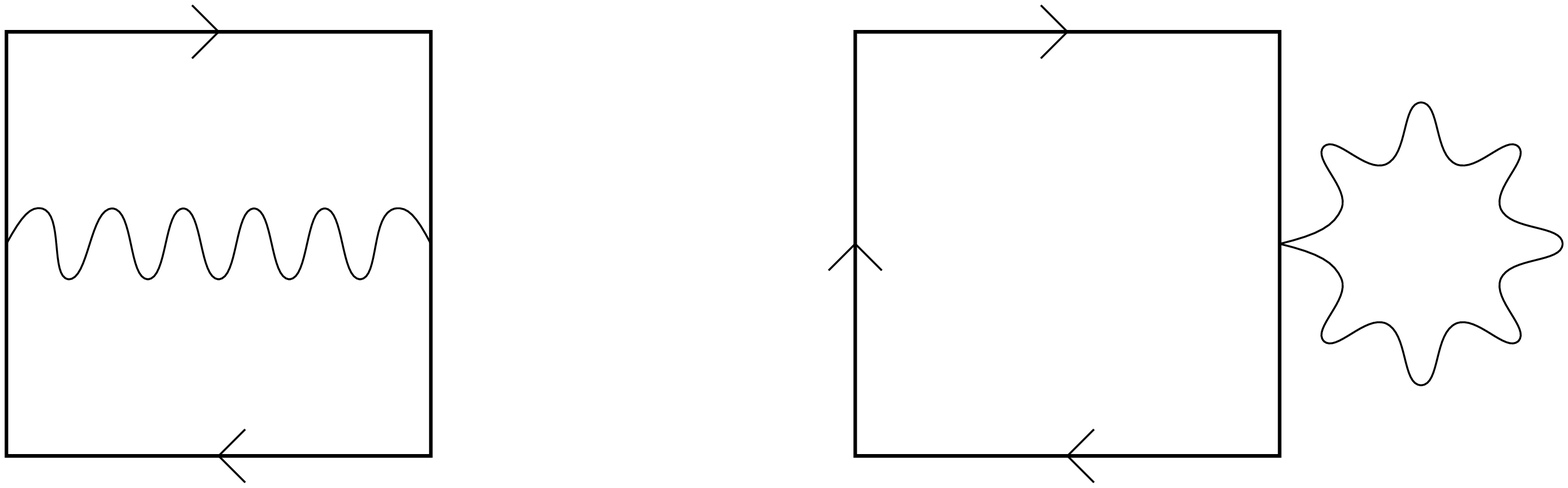,width=4in}}
\bigskip\bigskip
\centerline{\bf Figure 1}
\centerline{Photon corrections to the smallest loop excitation}
}
\vfill
\vbox{
\centerline{
\psfig{figure=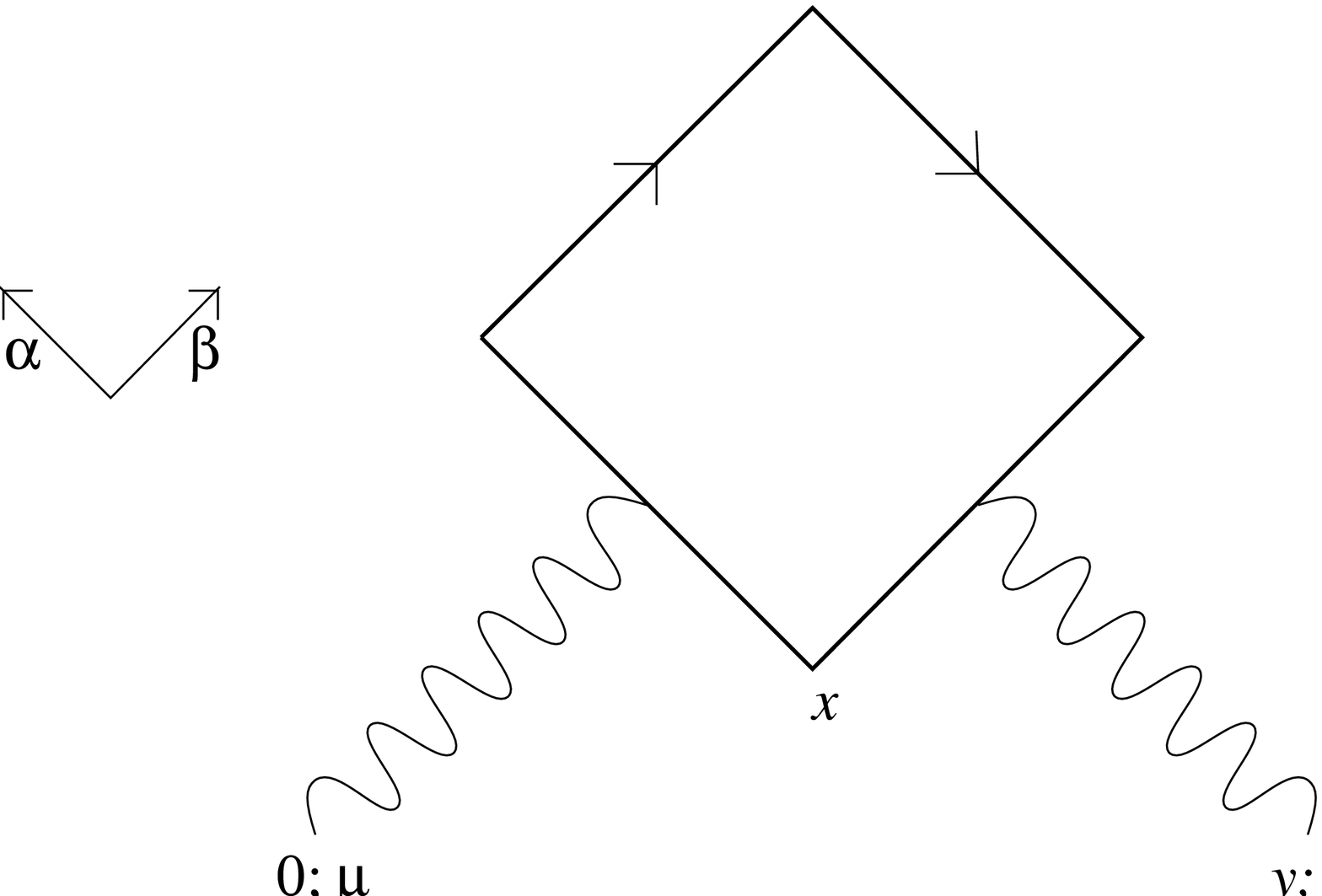,width=4in}}
\bigskip\bigskip
\centerline{\bf Figure 2}
\centerline{Loop correction to the photon propagator}
}
\end